\documentclass[sigconf,10pt,nonacm]{acmart}
\AtBeginDocument{%
  }

\usepackage{multirow}
\usepackage{algorithm}
\usepackage{algorithmic}
\usepackage{colortbl}
\usepackage{microtype}

\setcopyright{acmlicensed}
\copyrightyear{2018}
\acmYear{2018}
\acmDOI{XXXXXXX.XXXXXXX}
\acmConference[Conference acronym 'XX]{Make sure to enter the correct
  conference title from your rights confirmation emai}{June 03--05,
  2018}{Woodstock, NY}
\acmISBN{978-1-4503-XXXX-X/18/06}

\newcommand{\eg}{\textit{e.g., }}
\newcommand{\ie}{\textit{i.e., }}
\newcommand{\n}{\texttt{Prada}}

\begin{document}

\title{Prada: Black-Box LLM Adaptation with Private Data on Resource-Constrained Devices}

\author{Ziyao Wang$^{1}$, Yexiao He$^{1}$, Zheyu Shen$^{1}$, Yu Li$^{2}$, Guoheng Sun$^{1}$, \\Myungjin Lee$^{3}$, Ang Li$^{1}$}
\affiliation{%
  \institution{$^1$University of Maryland, College Park $\ $ $^2$University of California, Irvine $\ $ $^3$Cisco Research}
  \city{}
  \state{}
  \country{}
}
\email{{ziyaow,zyshen,yexiaohe,ghsun,angliece}@umd.edu, yul79@uci.edu, myungjle@cisco.com}


\begin{abstract}
In recent years, Large Language Models (LLMs) have demonstrated remarkable abilities in various natural language processing tasks. However, adapting these models to specialized domains using private datasets stored on resource-constrained edge devices, such as smartphones and personal computers, remains challenging due to significant privacy concerns and limited computational resources. Existing model adaptation methods either compromise data privacy by requiring data transmission or jeopardize model privacy by exposing proprietary LLM parameters.
To address these challenges, we propose {\n}, a novel privacy-preserving and efficient black-box LLM adaptation system using private on-device datasets. {\n} employs a lightweight proxy model fine-tuned with Low-Rank Adaptation (LoRA) locally on user devices. During inference, {\n} leverages the logits offset, \ie  difference in outputs between the base and adapted proxy models, to iteratively refine outputs from a remote black-box LLM. This offset-based adaptation approach preserves both data privacy and model privacy, as there is no need to share sensitive data or proprietary model parameters. Furthermore, we incorporate speculative decoding to further speed up the inference process of {\n}, making the system practically deployable on bandwidth-constrained edge devices, enabling a more practical deployment of {\n}.
Extensive experiments on various downstream tasks demonstrate that {\n} achieves performance comparable to centralized fine-tuning methods while significantly reducing computational overhead by up to 60\% and communication costs by up to 80\%.

\end{abstract}

%



\maketitle

\section{Introduction}
In recent years, Large Language Models (LLMs) have achieved remarkable performance across a wide range of natural language processing tasks. Their training paradigm typically involves pre-training on massive datasets, followed by supervised fine-tuning (SFT) for adaptation to specific downstream tasks. SFT has proven highly effective, significantly improving performance in domain-specific tasks  such as instruction following~\cite{peng2023instruction}, sentiment analysis~\cite{ding2024dynamicadaptiveoptimizationeffective}, medical report generation~\cite{lee2024llmcxrinstructionfinetunedllmcxr}, and code generation~\cite{li2024exploratorystudyfinetuninglarge}. 
Moreover, recent studies suggest that effective adaptation with SFT enhances reinforcement learning (RL) efficiency, improving model reasoning capabilities~\cite{guo2025deepseek}, making it a key research direction. 
However, the effectiveness of SFT in adapting LLMs relies heavily on task-specific datasets. Given that state-of-the-art (SOTA) LLMs have already been trained on nearly all publicly available text, researchers are increasingly exploring the potential of datasets stored on edge devices, such as smartphones, wearables, and medical equipment. When leveraged effectively, SFT on these on-device datasets can enable LLMs to specialize in advanced code generation, mathematical analysis, and medical image processing.

Beyond dataset availability, the widespread deployment of LLM applications and APIs on edge devices, such as smartphones and PCs, has further underscored the need for SFT LLM using on-device data for domain adaptation~\cite{qin2024enabling,mairittha2020improving}. 
Currently, users interact with LLMs through remote APIs that provide only generalized responses. However, this approach inherently lacks the capability to provide personalized LLM services tailored to users' specific needs. Since LLM developers cannot directly access on-device data, it becomes challenging to fine-tune models to individual users' preferences and usage contexts. As a result, there is a growing need to adapt LLMs with on-device datasets to enable personalized LLM services, making it a critical research direction in modern LLM development.

\textbf{Two-Sided Privacy Issues.}  
SFT LLMs using on-device data can be approached in two primary ways: (1) the \textit{data-sharing approach}~\cite{qin2024enabling}, which involves transmitting on-device data to LLM developers, \eg AI companies, for SFT on large-scale GPU clusters, or (2) the \textit{model-sharing approach}~\cite{mcmahan2017communication,wu2024multi}, which involves deploying the LLM to edge devices for local fine-tuning. However, both approaches present significant privacy risks, affecting either data privacy or model privacy.

The data-sharing approach raises serious privacy concerns. On-device datasets, particularly in sensitive domains such as healthcare and personal assistance, often contain highly sensitive information that cannot be transferred off-device~\cite{solangi2018future}. Users are understandably reluctant to share private data with third parties for fine-tuning purposes~\cite{kairouz2021advances, li2023privacy}. As a result, centralized SFT methods not only increase the risk of data leakage, but also discourage users from leveraging these approaches in privacy-sensitive applications.

While model-sharing eliminates data privacy concerns, it introduces risks to the proprietary parameters of the pre-trained LLM. Developing LLMs requires extensive training on vast datasets with significant computational resources, often involving thousands of GPUs over several months~\cite{grattafiori2024llama,guo2025deepseek,achiam2023gpt}. These models represent substantial investments in time, expertise, and capital, making their parameters valuable intellectual property. To protect these assets, many LLM developers opt to provide access only through APIs rather than sharing model parameters~\cite{achiam2023gpt}. As a result, deploying LLMs to edge devices for local fine-tuning is often infeasible, as developers are unwilling to expose their proprietary models. Even in cases where deployment is permitted, it creates risks of unauthorized access or unintended leakage~\cite{yao2024survey}, further discouraging this approach.

The dual challenge of protecting sensitive on-device data while safeguarding the proprietary nature of pre-trained LLMs highlights the complexity of enabling fine-tuning with on-device data. Addressing these challenges is essential to unlocking the potential of on-device datasets while ensuring both data privacy and model privacy.

\textbf{Limitations of Previous Solutions and Challenges.}
Various approaches have been proposed to protect data privacy and model privacy during SFT. One prominent direction involves using federated learning (FL) to leverage on-device data for SFT a global LLM~\cite{xia2021survey,wang2024flora, wang2025federated}. While FL mitigates data privacy concerns by keeping data on edge devices, it still requires the server to share LLM parameters with clients, sacrificing model privacy~\cite{niu2020billion}. Moreover, FL introduces significant computational and communication overhead on edge devices, which often have limited memory and bandwidth. These constraints make FL impractical for many real-world on-device SFT scenarios.
Another potential solution is to on-device fine-tune a local LLM with a private dataset and then employ knowledge distillation~\cite{gou2021knowledge}. However, due to the highly limited computational resources, training an on-device teacher LLM that surpasses the developer's base LLM in performance is nearly impossible, making it extremely difficult to perform effective knowledge distillation between the device and LLM developer~\cite{wu2023survey}. Recently, a class of methods termed Offsite-Tuning~\cite{xiao2023offsite,yao2024scaleot} has been proposed, which employs a transfer learning framework. In this approach, the LLM developer sends a lightweight adapter and a lossy compressed sub-model to the data owner. The adapter is then fine-tuned on private data locally with the assistance of the emulator. While this method resolves data privacy concerns, it only partially addresses model privacy. The device still receives a compressed version of the LLM, which reveals the backbone structure and part of model weights to users.

While these methods address parts of the privacy problem, existing approaches still struggle to effectively tackle both data privacy and model privacy in on-device dataset adaptation, especially under the computational and communication constraints of resource-limited edge devices. This remains an open challenge requiring further exploration.

\textbf{Our Solutions.}  
In this work, we propose {\n}, a privacy-preserving adaptation system for black-box LLMs that enables domain-specific fine-tuning using private on-device data, thereby addressing both data and model privacy challenges inherent in existing methods, all while operating efficiently within the resource constraints of edge devices. 
To achieve this, we propose a novel framework that leverages \textit{proxy model offsets} to adapt the black-box LLM. Specifically, we deploy a lightweight proxy model on the device, which has the same tokenizer as the black-box LLM. We fine-tune the proxy LLM on private datasets with LoRA to obtain the adapted proxy model. Then, during the inference, the device sends the user prompts to the black-box LLM for token generation, and gets back the token logits. The base proxy LLM and the adapted proxy LLM will generate the token logits on the device, and compute the logits offset between the two models, which represents the knowledge learned from the private dataset. Finally, the device uses the logits offset to iteratively prune the black-box token logits for adaptation until the end of generation. To minimize the inference latency, we also leverage the speculative decoding~\cite{leviathan2023fast} framework to fully utilize the KV cache~\cite{ge2023model} of LLMs, further speeding up the algorithm. Our proposed system tackles the two key challenges: (1) \textit{Two-Sides Privacy Preservation:} The edge device never accesses the black-box model parameters, and the server never directly sees the on-device private dataset, ensuring strong privacy guarantees; (2) \textit{Computational and Communication Efficiency:} Given the resource constraints of edge devices, {\n} minimizes both computational overhead and communication costs.

Our main contributions are as follows:
\begin{itemize}
    \item We propose {\n}, a novel system for fine-tuning black-box LLMs on-device with private data. {\n} enables privacy-preserving adaptation of black-box models while ensuring data confidentiality.
    \item Our framework is computationally efficient, making it suitable for resource-constrained edge devices. By leveraging a lightweight proxy model and parameter-efficient fine-tuning, {\n} significantly reduces the GPU memory usage by up to 80\%.
    \item {\n} incurs minimal communication costs, as it does not transmit LLM parameters or training datasets. This reduces the system communication cost by over 50\%, making it well-suited for edge computing scenarios.
    \item We conduct extensive experiments across multiple downstream tasks, demonstrating that {\n} achieves performance comparable to centralized fine-tuning while preserving both model and data privacy.
\end{itemize}

\section{Background and Motivation}
In this section, we first provide an overview of LLM fine-tuning and adaptation on edge devices. We then highlight the privacy concerns associated with existing adaptation methods and emphasize the critical necessity of developing {\n} for black-box LLM adaptation.

\subsection{LLM Adaptation with On-Device Data}
To leverage the valuable datasets available on edge devices for fine-tuning, there are two primary adaptation strategies: the \textit{data-sharing approach} and the \textit{model-sharing approach}. The data-sharing approach involves transmitting on-device data to LLM developers, who then perform fine-tuning on large-scale GPU clusters to adapt the model. Many LLM providers, such as OpenAI, collect user data to enhance their models' instruction-following capabilities and overall user experience~\cite{wang2023openchat}. In contrast, the model-sharing approach entails sharing the LLM architecture and parameter weights with the edge device, enabling local fine-tuning for domain-specific adaptation. Given the constrained computational capacity of edge devices and the substantial parameter size of LLMs, researchers have developed various parameter-efficient fine-tuning techniques, such as LoRA~\cite{hu2021lora}, BitFit~\cite{zaken2021bitfit}, and DoRA~\cite{liu2024dora}, alongside model compression methods like INT8 and FP8 quantization~\cite{wu2020integer,kuzmin2022fp8}, as well as model pruning strategies, including attention drop~\cite{he2024matters,yuan2024ditfastattn} and feed-forward network (FFN) drop~\cite{fan2019reducing,jaiswal2024ffn}. Furthermore, to mitigate privacy concerns in on-device LLM adaptation, researchers have explored FL for multi-device collaboration and Offsite-Tuning techniques that preserve data privacy while optimizing model performance.

\subsection{Model and Data Privacy Concerns Necessitate Black-Box Adaptation}
The motivation for adapting LLMs to on-device datasets is to enable personalized LLM services for specific domains. While both the data-sharing and model-sharing approaches facilitate LLM adaptation, they introduce critical privacy risks: the data-sharing approach compromises \textit{data privacy}, while the model-sharing approach jeopardizes \textit{model privacy}. 

The data-sharing approach requires edge devices to transmit their datasets—including potentially sensitive private information—to LLM developers. This issue has been extensively studied in the FL literature~\cite{mcmahan2017communication,wang2024flora,kuo2024federated,woisetschlager2024federated}, where methods have been proposed to mitigate privacy leakage. However, an equally significant but less explored concern arises in the model-sharing approach: \textit{model privacy}. Due to the high computational cost and complexity of training LLMs, as well as the fact that LLMs have become a key competitive advantage for technology companies, most LLM developers do not open-source their model parameters. Instead, they offer access to LLMs exclusively through APIs, thereby preserving proprietary model knowledge. This paradigm renders the model-sharing approach impractical, as it necessitates the release of LLM parameters to third-party edge devices, fundamentally compromising model privacy.

Given the pressing concerns of both data privacy and model privacy, there is an urgent need to develop an LLM adaptation framework that neither requires edge devices to share their training data nor mandates LLM developers to disclose model parameters. Designing such a system poses significant challenges due to the stringent privacy constraints and the limited computational resources available on edge devices. Addressing these challenges requires novel adaptation techniques that strike a balance between privacy preservation and efficient model adaptation in black-box settings.

\begin{figure}[t]
    \centering  \includegraphics[width=3.2in]{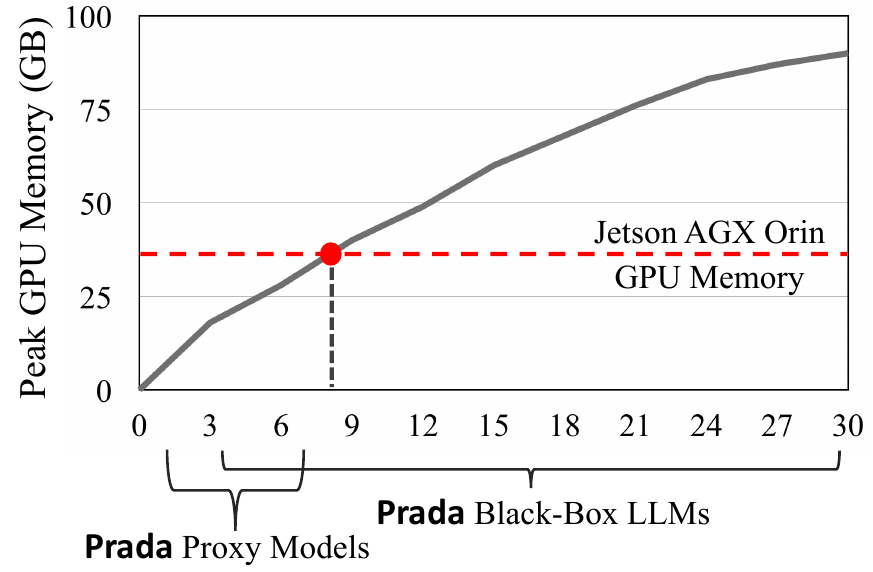}
    \caption{The relationship between Llama models parameter size and peak GPU memory usage during LoRA fine-tuning with rank=128. The red dashed line represents the GPU memory of the Jetson AGX device used in our experiments. Models above this threshold are difficult to fine-tune directly on-device. {\n} selects proxy models smaller than 7B, enabling on-device adaptation for black-box models as large as 30B.}
    \label{fig:cost}
\end{figure}

\subsection{Computational and Communication Constraints}

Beyond privacy concerns, black-box adaptation must also overcome the \textit{resource limitations} of edge devices. Unlike centralized servers, edge devices have constrained computational power, memory, and energy, making conventional fine-tuning infeasible. Previous model-sharing approaches, such as FL and offsite fine-tuning, have applied Parameter-Efficient Fine-Tuning (PEFT) methods like LoRA to reduce computational costs. However, the on-device computational overhead remains comparable to that of fine-tuning the full target model. For {\n}, minimizing the computational and memory requirements for fine-tuning black-box LLMs is a crucial optimization objective. As shown in Figure \ref{fig:cost}, most of the black-box LLMs we aim to adapt in {\n} exceed the on-device GPU memory limit. Therefore, {\n} needs to use proxy models within the GPU capacity (\ie less than 8B parameter size) for the on-device fine-tuning. In addition, transmitting LLM parameters is also a huge overhead for edge devices. Therefore, sending and receiving only prompts, tokens, and logits rather than the LLM parameters is also a crucial optimization target of {\n}.

Table \ref{tab:limitations} shows the comparison between current approaches and {\n} in terms of four main properties within the scope of on-device data SFT. These properties include: (1) model privacy: the ability to ensure that the pre-trained LLM parameters remain inaccessible to edge devices; (2) data privacy: the ability to prevent the server from accessing on-device training datasets; (3) memory efficiency: fine-tune the model with reduced computational overhead compared to standard PEFT fine-tuning; and (4) communication efficiency: the ability to avoid transmitting LLM weights, thereby minimizing communication overhead. 
As shown in the table, only {\n} satisfies all four properties, demonstrating its effectiveness in enabling privacy-preserving and resource-efficient adaptation for black-box LLMs on edge devices. This highlights {\n} as a superior solution compared to existing approaches, making it a viable choice for real-world deployment in privacy-sensitive and resource-constrained environments.

\begin{table}[t]
\caption{Comparison of LLM adaptation methods with {\n} in terms of model privacy, data privacy, memory consumption, and communication cost. Two black squares indicate that the corresponding method excels in the given aspect. A black and a white square signify partial effectiveness, but not fully achieved. Two white squares indicate that the method has no effect in the corresponding aspect.}
\centering
\resizebox{8.3cm}{!}{
\begin{tabular}{c|cccc}
\toprule
\multirow{2}{*}{\centering \textbf{Method}} & \multicolumn{1}{c}{\textbf{Model}} & \multicolumn{1}{c}{\textbf{Data}} & \multicolumn{1}{c}{\textbf{Memory}} & \multicolumn{1}{c}{\textbf{Comm.}} \\  
 & \multicolumn{1}{c}{\textbf{Privacy}} & \multicolumn{1}{c}{\textbf{Privacy}} & \multicolumn{1}{c}{\textbf{Efficiency}} & \multicolumn{1}{c}{\textbf{Efficiency}} \\
\midrule
\textbf{Data-Sharing} & $\blacksquare \blacksquare$ & $\square \square$ & $\square \square$ & $\square \square$ \\
\textbf{Distillation}  & $\blacksquare \blacksquare$ & $\blacksquare \blacksquare$ & $\square \square$ & $\square \square$ \\ 
\textbf{Fed. Learning}  & $\square \square$ & $\blacksquare \blacksquare$ & $\square \square$ & $\square \square$ \\
\textbf{Offsite-Tuning}  & $\blacksquare \square$ & $\blacksquare \blacksquare$ & $\blacksquare \square$ & $\square \square$ \\
\midrule
\textbf{{\n}} & $\blacksquare \blacksquare$ & $\blacksquare \blacksquare$ & $\blacksquare \blacksquare$ & $\blacksquare \blacksquare$ \\  
\bottomrule
\end{tabular}}
\label{tab:limitations}
\end{table}

\section{System Overview}
In this work, we design {\n}, a black-box LLM adaptation framework that enables on-device fine-tuning using private data. In {\n}, edge devices adapt a black-box LLM without sharing private datasets or accessing the model parameters. With the deployment of {\n}, the LLM developer (denoted as the "server" in the following sections) can facilitate LLM adaptation for serving edge devices (denoted as "clients"), while only exchanging prompts and model output between the server and clients.

\begin{figure*}[t]
    \centering  \includegraphics[width=.9\textwidth]{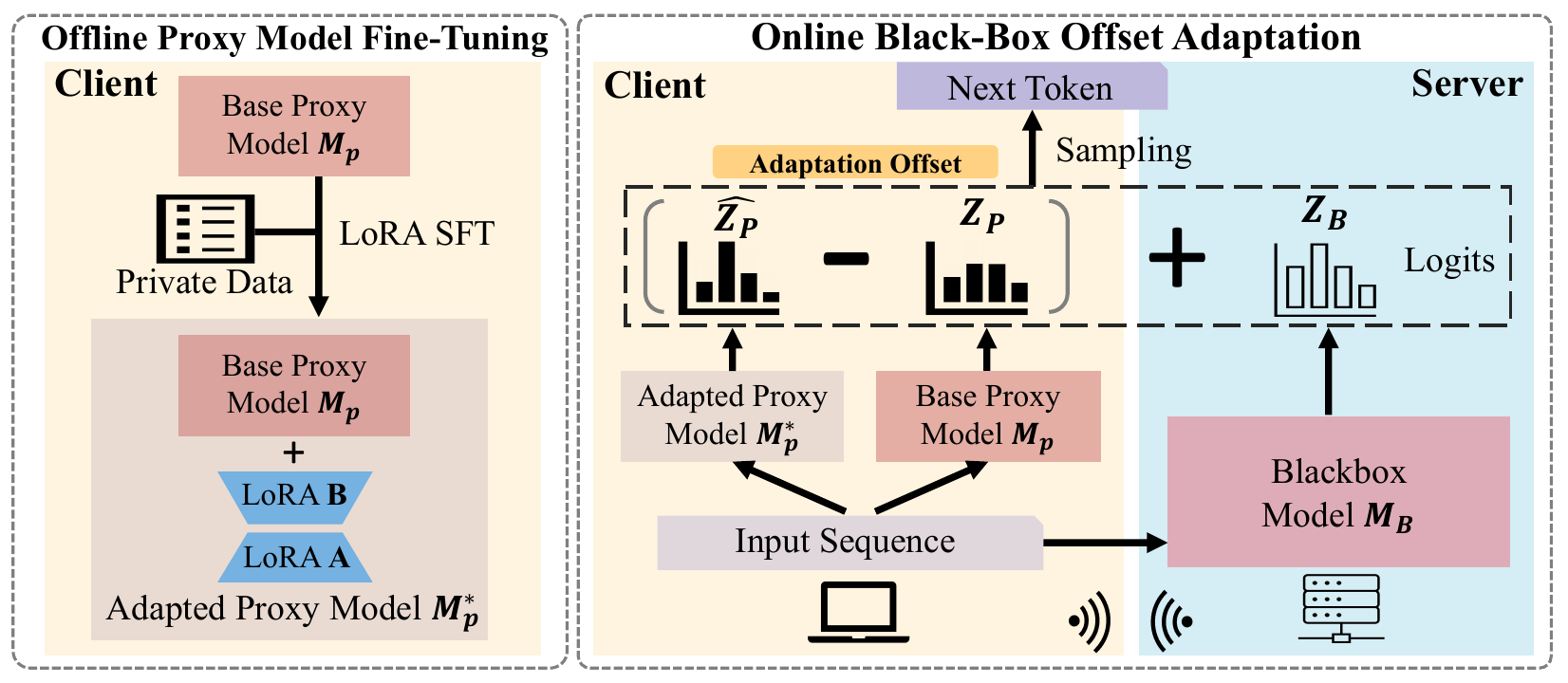}
    \caption{The overview of {\n}. {\n} begins with the client offline fine-tuning the proxy model using LoRA. Subsequently, the client engages in online black-box offset adaptation through interactions with the server.}
    \label{fig:overview}
\end{figure*}

Figure \ref{fig:overview} illustrates the architecture and workflow of the proposed framework. {\n} consists of two main steps. The first step is the offline proxy model fine-tuning, where the client selects a proxy LLM and applies LoRA fine-tuning on its local private dataset using local computational resources. 
The second step is the online black-box offset adaptation. In this step, the client performs inference by interacting with the server. The client first sends the user’s prompt to the server, which then queries the black-box LLM to generate the next token logits. The server transmits these logits back to the client. Simultaneously, the client conducts two forward passes using both the base proxy model and the adapted proxy model with the same input as sent to the server. The offset (difference) between the adapted proxy model’s logits and the base proxy model’s logits is computed and added to the received black-box LLM logits to refine the final token prediction. The client then returns the adjusted token to the server, which proceeds with autoregressive generation until completion.
The black-box offset adaptation can be optimized by speculative decoding, enabling the server to generate multiple tokens in one iteration, while the client simultaneously refines these tokens in batches.
By leveraging this offset-based correction, {\n} effectively adapts the black-box LLM to local data while preserving both data privacy and model privacy. This approach minimizes reliance on the server while ensuring domain-specific adaptation without direct access to the LLM’s parameters.

\section{System Design}

In this section, we describe the key components of {\n}, including the fine-tuning of the proxy model using LoRA and the offset-based adaptation mechanism for refining black-box LLM predictions. We start from defining the problem settings and then dive into the details of each component.

\subsection{Problem Settings}
We consider a server-client collaboration scenario where the server hosts a black-box LLM, denoted as $\mathcal{M}_B$, while the client fine-tunes a lightweight base proxy model, $\mathcal{M}_P$, on the local private dataset $\mathcal{D}$. This fine-tuning process yields an adapted proxy model, $\widehat{\mathcal{M}}_p$, which incorporates knowledge from the local dataset. $\mathcal{M}_B$ and $\mathcal{M}_P$ need to have the same tokenizer for logits dimension alignment and token correspondence.
During inference, the client initiates a query by providing a prompt token sequence $\{x_0, \dots, x_k\}$. The client and server engage in an iterative communication process and ultimately produce a response sequence $\{x_{k+1}, \dots, x_{k+t}\}$. 
This collaborative paradigm enables the client to leverage the powerful capabilities of the black-box LLM while ensuring that the generated responses are aligned with the characteristics of the local dataset $\mathcal{D}$.

\subsection{Offline Proxy Model LoRA Fine-Tuning}
To enable efficient adaptation, the client selects a lightweight \textit{proxy model} and applies \textit{Low-Rank Adaptation} (LoRA) for fine-tuning on the local dataset. Given the computational constraints of edge devices, full fine-tuning of a large model is impractical. Instead, LoRA introduces trainable low-rank matrices into the transformer's weight matrices while keeping the pre-trained model parameters frozen, significantly reducing the number of trainable parameters.

Let $\mathcal{M}_P$ denote the base proxy model with weight matrix $\mathcal{W}_p^{(i)} \in \mathbb{R}^{m \times n}$ for the $i$-th transformer layer. Instead of directly updating $\mathcal{W}_p^{(i)}$, LoRA injects trainable low-rank matrices $A^{(i)}$ and $B^{(i)}$ into each layer, modifying the weights as follows:

\begin{equation}
\widehat{\mathcal{W}}_p^{(i)} = \mathcal{W}_p^{(i)} + B^{(i)} A^{(i)},
\end{equation}
where $A^{(i)}, B^{(i)}$ are two low-rank matrices with $A^{(i)} \in \mathbb{R}^{m \times r}$ and $B^{(i)} \in \mathbb{R}^{r \times n}$. The rank $r$ is much smaller than the original weight matrix dimensions $m$ and $n$, i.e., $r \ll \min(m, n)$, thus allowing efficient adaptation with significantly fewer trainable parameters compared to full fine-tuning.
During local fine-tuning, the weights of the base proxy model are frozen. The LoRA weights will be updated to adapt the model to the local dataset:

\begin{equation}
(A, B) = \arg\min_{A, B} \sum_{(x, y) \in \mathcal{D}} \mathcal{L} \big( \widehat{\mathcal{M}}_p(x), y \big).
\end{equation}

After training, we store all LoRA module weights alongside the base proxy model weights $\mathcal{M}_P$ without merging them. This design enables efficient inference: when using $\mathcal{M}_P$, we can directly load its pre-trained parameters, while for inference with $\widehat{\mathcal{M}}_P$, the LoRA parameters are loaded on top of the base model, preserving modularity and flexibility.

By selecting a lightweight proxy model for LoRA fine-tuning, we achieve the following benefits:
(1) Reduced memory usage during fine-tuning stage by using a lightweight proxy model,
(2) further decreased computational overhead with LoRA, and (3) optimized storage and inference memory consumption by leveraging the base model with LoRA saving and loading mechanism.

\begin{figure}[t]
    \centering  \includegraphics[width=3.3in]{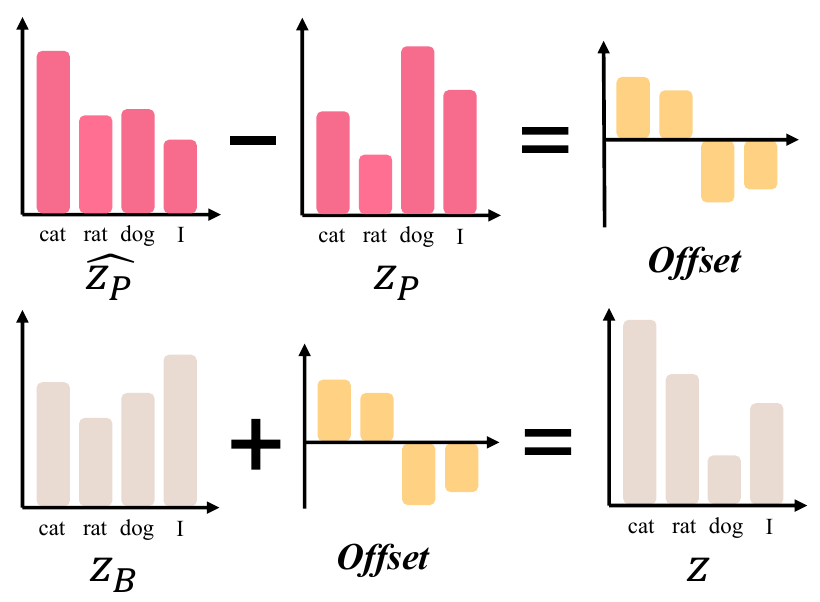}
    \caption{The offset adaptation of {\n}.}
    \label{fig:offset}
\end{figure}

\subsection{Online Black-Box Offset Adaptation}
\label{sec:proxy}
After fine-tuning the proxy model, we obtain an adapted proxy model $\widehat{\mathcal{M}}_P$ and a base proxy model $\mathcal{M}_P$ on the client, and a black-box LLM $\mathcal{M}_B$ on the server. The inference process follows an autoregressive generation framework, involving iterative interactions between the client and the server. At each iteration $t$, the process unfolds as follows:

\paragraph{Black-Box LLM Inference.} If $t=1$, the client sends the full input token sequence $\{x_0, \dots, x_k\}$ to the server. Otherwise, the client transmits only the latest generated token $x_{k+t-1}$ in the last iteration. The server then constructs the input sequence by concatenating this new token with the previous tokens and performs the next-token prediction:

\begin{equation}
z_B^{(t)} = \mathcal{M}_B(x_1, \dots, x_{k+t-1}),
\end{equation}
where $z_B^{(t)}$ represents the next-token logits produced by the black-box LLM. The server then returns $z_B^{(t)}$ to the client.

\paragraph{Proxy LLM Inference.} 
Using the same input sequence, both the base proxy model and the adapted proxy model generate next-token logits:

\begin{equation}
\begin{split}
z_P^{(t)} &= \mathcal{M}_P(x_1, \dots, x_{k+t-1}), \\
\widehat{z}_P^{(t)} &= \widehat{\mathcal{M}}_P(x_1, \dots, x_{k+t-1}).
\end{split}
\end{equation}

Since the proxy models and the black-box LLM share the same tokenizer, the logits $z_P^{(t)}$, $\widehat{z}_P^{(t)}$, and $z_B^{(t)}$ have identical dimensions and token-wise correspondence.

\paragraph{Offset Logits Adjustment.} 
The adaptation of $\widehat{\mathcal{M}}_P$ from $\mathcal{M}_P$ leads to differences in their predicted logits with the same input, reflecting the knowledge acquired from local fine-tuning. Consequently, we define the difference between $\widehat{z}_P^{(t)}$ and $z_P^{(t)}$ as the \textit{adaptation offset}, capturing token-wise probability shifts due to local adaptation. This offset is applied to the server logits $z_B^{(t)}$ to incorporate local knowledge into the black-box model’s generation:

\begin{equation}
\label{eq:offset}
z^{(t)} = z_B^{(t)} + \underbrace{\widehat{z}_P^{(t)} - z_P^{(t)}}_{\text{Adaptation Offset}}.
\end{equation}

Figure \ref{fig:offset} intuitively illustrates the process of generating the offset and applying it to $z_B^{(t)}$, which essentially prunes the black-box model's output by leveraging the logits difference induced by local fine-tuning. Finally, the next token is sampled from the adjusted logits:

\begin{equation}
x_{k+t} \sim z^{(t)}.
\end{equation}

$x_{k+t}$ is then appended to the input sequence for the next iteration, continuing the autoregressive process.

\subsection{Inference Speedup through Speculative Decoding}
The per-token generation process described in Section~\ref{sec:proxy} mathematically outlines how {\n} operates. However, for long-sequence generation tasks, this approach may be inefficient due to the lack of KV cache utilization and other sequence generation optimizations. This inefficiency highlights the need for algorithm-level improvements to better leverage existing inference optimization techniques.

To address this, we adopt the draft-then-verify paradigm of speculative decoding to accelerate {\n}. Specifically, in each step of the black-box offset adaptation process, we apply the following modifications:

\paragraph{Black-Box LLM Inference - $S$-token Sequence Generation.} Rather than predicting only the next token, the black-box LLM $\mathcal{M}_B$ generates a token sequence with predefined length $S$:
\begin{equation}
    \{ x_{k+i}, z_B^{(i)} \}_{i=1}^{S} = \mathcal{M}_B(x_1, \dots, x_{k+i-1}),
\end{equation}
where $x_{k+i}$ denotes the generated token, and $z_B^{(i)}$ represents its corresponding logits. This sequence can be generated by applying the KV cache and Flash Attention~\cite{dao2022flashattention}. The server then collects the generated sequence $\{x_{k+1}, \dots, x_{k+S}\}$ with $S$ new tokens and the associated logits $\{z_B^{(1)}, \dots, z_B^{(S)}\}$, sending them to the client.

\paragraph{Proxy LLMs Inference - Generation in Parallel.} Given the received $S$ tokens, the base proxy model and the adapted proxy model generate logits in parallel:
\begin{equation}
\begin{aligned}
    z_{P}^{(i)} &= \mathcal{M}_P(x_1, \dots, x_{k+i}), \\
    \widehat{z}_{P}^{(i)} &= \hat{\mathcal{M}}_P(x_1, \dots, x_{k+i}), \quad \forall i \in \{1, \dots, S-1\}.
\end{aligned}
\end{equation}

The client generation process here will be significantly faster since all token logits can be computed in parallel, eliminating the need for sequential decoding.

\paragraph{Offset Logits Adjustment - Verify, Accept, and Drop.} The client iterates through the tokens, applying offset adaptation from Eq.~\ref{eq:offset} to adjust the logits received from the server. The first token modified by offset adaptation (\ie $z^{(i)} \neq z_B^{(i)}$) determines the acceptance boundary:
\begin{itemize}
    \item Accept all tokens preceding the modified token and concatenate them with the previous input sequence.
    \item Replace the modified token with its offset-adapted counterpart.
    \item Discard all subsequent tokens.
\end{itemize}
This new input sequence is sent to the server for the next iteration. The iteration continues until the EOS token is accepted by the client. The detailed algorithm for this speculative-decoding-enhanced {\n} is presented in Algorithm \ref{alg:spec}.

\begin{algorithm}[h]
\caption{Speculative-Decoding-Enhanced Black-Box Offset Adaptation ({\n}-SD)}
\label{alg:spec}
\renewcommand{\algorithmicrequire}{\textbf{Input:}}
\renewcommand{\algorithmicensure}{\textbf{Output:}}
\begin{algorithmic}[1]
\REQUIRE Input sequence $\{x_1, \dots, x_k\}$, black-box LLM $\mathcal{M}_B$, base proxy model $\mathcal{M}_P$, adapted proxy model $\hat{\mathcal{M}}_P$;
\ENSURE Offset adapted Black-Box LLM generation;
\WHILE{EOS token is not accepted}
    \FOR{$i = 1$ to $S$}
        \STATE $z_B^{(i)} = \mathcal{M}_B(x_1, \dots, x_{k+i-1}),$
        \STATE $\dot{x}_{k+i} \sim z_B^{(i)}$
    \ENDFOR
    \STATE Client receives tokens and logits
    \STATE $i = 1, \dots, S \ $ \textbf{in parallel do}
    \STATE $ \quad z_{P}^{(i)} = \mathcal{M}_P({x}_1, \dots, x_{k+i}),$
    \STATE $ \quad \widehat{z}_{P}^{(i)} = \widehat{\mathcal{M}}_P({x}_1, \dots, x_{k+i}),$
    \FOR{$i = 1$ to $S$}
        \STATE $z^{(i)} = z_B^{(i)} + \widehat{z}_P^{(i)} - z_P^{(i)},$
        \STATE $x_{k+i} \sim z^{(i)},$
        \IF{$x_{k+i} \neq \dot{x}_{k+i}$}
            \STATE Append $\{x_{k+1}, \dots, x_{k+i}\}$ to the input
            \STATE Discard $\{\dot{x}_{k+i+1}, \dots, \dot{x}_{k+S}\}$
            \STATE Break loop
        \ENDIF
    \ENDFOR
    \STATE Update $k \leftarrow k+i$ 
    \STATE Send new input sequence to server for next iteration
\ENDWHILE
\RETURN Final generated sequence
\end{algorithmic}
\end{algorithm}

\section{Experiments}
\subsection{Experimental Setup}
\textbf{Models.} We evaluate {\n} using three popular LLM families: Llama~\cite{touvron2023llama}, Qwen~\cite{qwen2.5,qwen2}, and Yi~\cite{ai2024yi}. The models used in our experiments are listed in Table \ref{tab:models}. For Llama models, we use Llama-7B and Llama-2-7B as proxy models and larger Llama-13B, Llama-2-13B, and Llama-30B as black-box LLMs to simulate the server-client collaboration. For Qwen models, Qwen2-1.5B and Qwen2.5-1.5B serve as proxy models, while Qwen2.5-3B, Qwen2-7B, and Qwen2.5-7B act as black-box LLMs. For the Yi family, we use two 6B models as proxy models and 9B and 34B models as black-box LLMs. Our proxy models are capped at 7B parameters, ensuring compatibility with the 32GB GPU memory of edge device, effectively demonstrating {\n}'s efficiency and practicality. The models specified in Table \ref{tab:models} (\eg Black-Box Model 1) will be directly referenced in subsequent experiments without repeating their full names.

\begin{table}[t]
\caption{Models used in experiments.}
\centering
\resizebox{8.5cm}{!}{
\begin{tabular}{c|ccc|cc}
\toprule
\multirow{2}{*}{\textbf{Family}} & \multicolumn{3}{c|}{\textbf{Black-Box Models}} & \multicolumn{2}{c}{\textbf{Proxy Models}} \\  
 & \textbf{Model 1} & \textbf{Model 2} & \textbf{Model 3} & \textbf{Model 1} & \textbf{Model 2} \\  
\midrule
\textbf{Llama}  & Llama-30B & Llama2-13B & Llama-13B & Llama2-7B & Llama-7B \\  
\textbf{Qwen}  & Qwen2.5-7B & Qwen2-7B & Qwen2.5-3B & Qwen2.5-1.5B & Qwen2-1.5B \\  
\textbf{Yi}  & Yi-34B & Yi1.5-9B & Yi-9B & Yi1.5-6B & Yi-6B \\ 
\bottomrule
\end{tabular}}
\label{tab:models}
\end{table}

\begin{table*}[t]
  \caption{Performance of black-box adaptation on Llama, Qwen, and Yi models. Rows highlighted with a light gray background represent our {\n} approach with different proxy models.}
  \vspace{-0.3cm}
  \label{tab:main}
  \centering
  \resizebox{17cm}{!}{
   \begin{tabular}{c|ccc|ccc|ccc}
    \toprule
     \multirow{2}{*}{\textbf{Methods}} 
     &\multicolumn{3}{c|}{\textbf{Llama-30B}} 
     &\multicolumn{3}{c|}{\textbf{Llama2-13B}} 
     &\multicolumn{3}{c}{\textbf{Llama-13B}} \\
    \cmidrule(r){2-4} \cmidrule(r){5-7} \cmidrule(r){8-10}
    &\textbf{MMLU}&\textbf{HumanEval}&\textbf{GSM8K} 
    &\textbf{MMLU}&\textbf{HumanEval}&\textbf{GSM8K} 
    &\textbf{MMLU}&\textbf{HumanEval}&\textbf{GSM8K} \\
    \midrule
    \textbf{Base} 
    &60.70&20.12&25.33 &59.92&18.29&15.77 &43.58&15.24&12.81 \\
    \midrule
    \textbf{Black-Box LLM SFT}
    &61.10&\textbf{26.83}&26.68 &62.18&23.17&\textbf{26.30} &60.06&23.17&21.83 \\
    \textbf{Offsite-Tuning}
    &55.25&19.86&25.01 &60.18&20.95&21.34 &50.15&18.29&16.32 \\
    \rowcolor{gray!15} \textbf{{\n}-Llama-7B} &
    56.26&26.22&27.37 &55.84&\textbf{24.39}&21.99 &51.09&\textbf{23.78}&16.52 \\
    \rowcolor{gray!15} \textbf{{\n}-Llama2-7B} &
    \textbf{65.86}&25.00&\textbf{27.45} &\textbf{68.00}&21.34&20.55 &\textbf{64.57}&20.12&\textbf{24.26} \\
    \midrule \midrule
     \multirow{2}{*}{\textbf{Methods}} 
     &\multicolumn{3}{c|}{\textbf{Qwen2.5-7B}} 
     &\multicolumn{3}{c|}{\textbf{Qwen2-7B}} 
     &\multicolumn{3}{c}{\textbf{Qwen2.5-3B}} \\
    \cmidrule(r){2-4} \cmidrule(r){5-7} \cmidrule(r){8-10}
    &\textbf{MMLU}&\textbf{HumanEval}&\textbf{GSM8K} 
    &\textbf{MMLU}&\textbf{HumanEval}&\textbf{GSM8K} 
    &\textbf{MMLU}&\textbf{HumanEval}&\textbf{GSM8K} \\
    \midrule
    \textbf{Base} 
    &73.86&50.00&77.37 &72.15&45.73&77.86 &65.32&34.15&64.22 \\
    \midrule
    \textbf{Black-Box LLM SFT} 
    &71.03&\textbf{55.49}&\textbf{82.41} &73.00&43.90&77.41 &\textbf{71.71}&\textbf{38.41}&75.82 \\
    \textbf{Offsite-Tuning}
    &70.60&51.22&74.14 &73.14&43.90&76.35 &64.69&34.75&70.19 \\
    \rowcolor{gray!15} \textbf{{\n}-Qwen2-1.5B} &
    \textbf{74.45}&51.83&78.31 &\textbf{74.53}&45.12&78.54 &69.10&34.75&\textbf{77.18} \\
    \rowcolor{gray!15} \textbf{{\n}-Qwen2.5-1.5B} &
    72.84&52.44&81.24 &74.11&\textbf{48.78}&\textbf{78.81} &69.34&35.37&74.75 \\
    \midrule \midrule
     \multirow{2}{*}{\textbf{Methods}} 
     &\multicolumn{3}{c|}{\textbf{Yi-34B}} 
     &\multicolumn{3}{c|}{\textbf{Yi-1.5-9B}} 
     &\multicolumn{3}{c}{\textbf{Yi-9B}} \\
    \cmidrule(r){2-4} \cmidrule(r){5-7} \cmidrule(r){8-10}
    &\textbf{MMLU}&\textbf{HumanEval}&\textbf{GSM8K} 
    &\textbf{MMLU}&\textbf{HumanEval}&\textbf{GSM8K} 
    &\textbf{MMLU}&\textbf{HumanEval}&\textbf{GSM8K} \\
    \midrule
    \textbf{Base} 
    &76.13&48.17&69.87 &60.55&38.41&65.58 &61.80&36.59&55.95 \\
    \midrule
    \textbf{Black-Box LLM SFT}
    &\textbf{78.04}&\textbf{54.27}&72.85 &\textbf{69.63}&\textbf{44.17}&\textbf{75.42} &\textbf{71.07}&\textbf{42.08}&64.34 \\
    \textbf{Offsite-Tuning} 
    &74.74&51.83&75.85 &66.61&42.25&72.14 &67.98&40.25&61.55 \\
    \rowcolor{gray!15} \textbf{{\n}-Yi-6B} 
    &76.13&52.44&73.19 &60.56&43.90&72.18 &61.80&39.63&59.06 \\
    \rowcolor{gray!15} \textbf{{\n}-Yi1.5-6B} 
    &77.65&53.66&\textbf{80.55} &69.58&41.46&68.73 &63.28&39.63&\textbf{65.95} \\
    \bottomrule
  \end{tabular}}
\end{table*}

\textbf{Tasks, Datasets, and Benchmarks.} We evaluate the adaptation performance of {\n} on three representative downstream tasks: question answering (QA), mathematical problem-solving, and code generation. For QA, we fine-tune the proxy model on a selected MMLU-10K~\cite{he2024shed} training dataset and evaluate offset adaptation on the MMLU~\cite{hendrycks2020measuring} benchmark. For mathematical problem-solving, we fine-tune the proxy model on  MetaMathQA~\cite{yu2023metamath} dataset and evaluate on  GSM8K~\cite{cobbe2021training} benchmark. For code generation, we fine-tune the proxy model on  Code-Instruction-120K dataset and measure performance using the HumanEval~\cite{chen2021evaluating} pass@1 metric.


\textbf{System Implementation.}
To demonstrate that {\n} can successfully operate on edge devices, experiments for {\n} are conducted on a Jetson AGX Orin device, which has a memory capacity of 32 GB, representing a spectrum of resource-constrained environments. Our results show that {\n} can be fully executed on this device, whereas baseline approaches, such as directly fine-tuning the black-box LLM, are infeasible due to GPU memory limitations. This highlights {\n}'s ability to efficiently operate on edge devices while meeting the computational and communication efficiency requirements we defined.

For baseline experiments that cannot be executed on edge devices, we use a GPU cluster with 8 NVIDIA A6000 GPUs.

\textbf{Baselines.} We evaluate the adaptation performance, the fine-tuning stage cost, and the inference stage cost of {\n} by comparing it against the following baselines:
\begin{itemize}
    \item \textbf{Black-Box LLM SFT:} A data-sharing approach that transfers the dataset to the LLM developer and directly fine-tunes the black-box LLM. We use the same hyperparameter settings as our proxy model fine-tuning for the black-box LLM.
    \item \textbf{Offsite-Tuning~\cite{xiao2023offsite,frikha2024obfuscatune}:} A privacy-preserving adaptation approach that fine-tunes adapters of a compressed version of the black-box LLM. We follow the default settings in the Offsite-Tuning paper and adapt the model to our black-box LLMs.
\end{itemize}

These baselines do not ensure either data or model privacy.

\textbf{Hyperparameters.} For the local LoRA fine-tuning, we set the LoRA rank to 128 and apply LoRA to both the attention layers and the FFN layers. We set the learning rate to 3e-4 and the batch size to 4. For the MMLU-10k dataset, we fine-tune the proxy model for 3 epochs. For MetaMathQA and Code-Instruction-120K, we only fine-tune the model for 1 epoch. For the black-box model fine-tuning in the baselines, we adopt the same hyperparameters. For the Offsite-Tuning experiments, we use adapters on 8 transformer layers and adopt the same hyperparameter settings of their paper~\cite{xiao2023offsite}. For speculative decoding experiments, we use $S=8$.

\subsection{Black-Box Adaptation Performance}
We compare the performance of {\n} with baselines across three model families and three benchmarks. The main results are summarized in Table \ref{tab:main}.

\textbf{Llama Model Family.} {\n} demonstrates strong adaptation performance across all black-box models. For the two proxy models, Llama-7B and Llama-2-7B, {\n} consistently outperforms Offsite-Tuning across all the tasks. For example, on Llama-13B, {\n}-Llama-7B achieves 51.09\% on MMLU, 23.78\% on HumanEval, and 16.52\% on GSM8K, exceeding Offsite-Tuning (50.15\%, 18.29\%, and 16.32\% on three benchmarks, respectively) by reducing the gap with fine-tuning the black-box LLM. Similarly, on Llama-30B, {\n}-Llama-7B improves the performance of all three benchmarks compared to Offsite-Tuning, especially outperforms 6.36\% on HumanEval, demonstrating its ability to improve reasoning tasks even without accessing the parameters of the black-box model.

Interestingly, {\n}-Llama-2-7B even outperforms the performance of directly fine-tuned larger models in certain cases. Specifically, {\n}-Llama-2-7B achieves MMLU scores of 64.57\%, 65.86\%, and 68.00\%, surpassing the directly fine-tuned Llama-13B model (60.06\%), Llama-30B (61.10\%), and Llama-2-13B (62.18\%). This suggests that offset-based adaptation, when paired with a strong proxy model, can achieve better generalization than directly fine-tuning the large black-box model, potentially due to the structured knowledge captured by proxy fine-tuning. 

\textbf{Qwen Model Family.} {\n} continues to perform robustly across different model sizes. On Qwen2.5-3B, {\n} with a Qwen2-1.5B proxy model achieves a HumanEval score of 34.75\% and a GSM8K score of 77.18\%, surpassing fully supervised fine-tuning. Compared to Offsite-Tuning, which struggles to maintain performance in mathematical reasoning and coding tasks, {\n} consistently demonstrates superior adaptation quality. On Qwen2-7B, {\n} further achieves an MMLU score of 74.11\%, reinforcing the observation that the offset-based approach enables effective knowledge transfer from the proxy model to the black-box LLM without direct parameter updates. 

\textbf{Yi Model Family.} Similar trends hold true for the Yi family models, the results highlight the generalization capability of {\n} across different model architectures, showing that its effectiveness is not restricted to a specific LLM family.

Across all three model families and benchmarks, {\n} consistently demonstrates competitive or superior performance compared to Offsite-Tuning and even matches or surpasses centralized fine-tuning methods in certain scenarios, particularly in larger black-box LLMs where adaptation is more challenging. 
The effectiveness of {\n} is positively correlated with the proxy model quality, suggesting that selecting a lightweight yet advanced proxy base model significantly benefits the performance of {\n}. The experiment results that {\n} achieves comparable results to direct fine-tuning while ensuring the data and model privacy is very promising.
These findings highlight the potential of {\n} as a practical and scalable solution for black-box LLM adaptation in privacy-sensitive environments.

\begin{figure}[t]
    \centering  \includegraphics[width=3.4in]{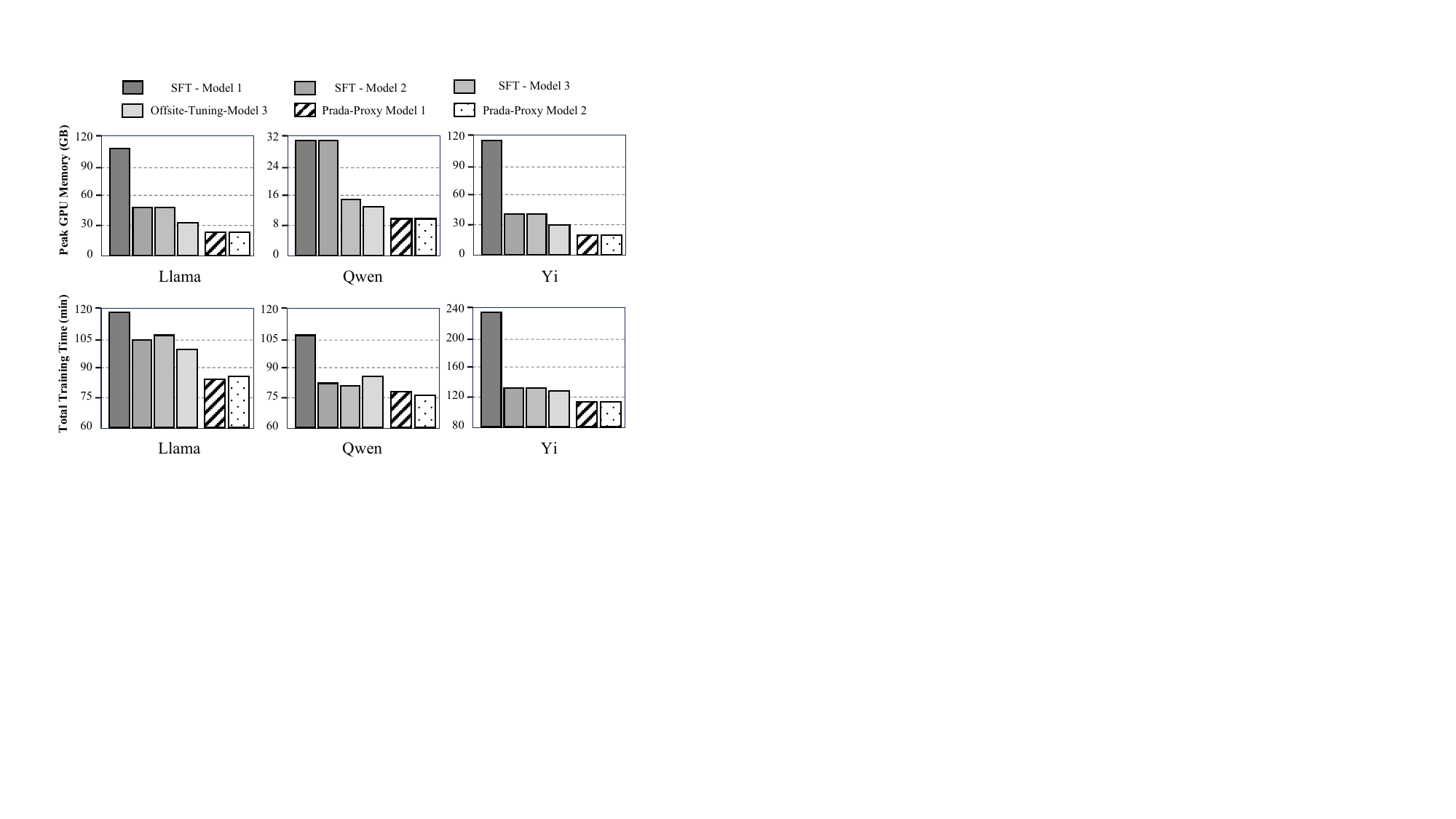}
    \vspace{-0.3cm}
    \caption{Comparison of the single-GPU peak memory usage (top row) and total training time (bottom row) on MMLU training dataset of {\n} and baselines during the SFT stage. The memory of Llama-30B and Qwen-34B are estimated.}
    \label{fig:ftcost}
\end{figure}

\begin{table*}[t]
  \caption{Comparison of {\n} and baselines in average token-wise latency (\textbf{ms/token}) on Qwen models. The average token-wise latency is calculated by the total running time divided by the number of total response tokens.}
  \vspace{-0.3cm}
  \label{tab:latency}
  \centering
  \resizebox{17cm}{!}{
   \begin{tabular}{c|ccc|ccc|ccc}
    \toprule
     \multirow{2}{*}{\textbf{Methods}} 
     &\multicolumn{3}{c|}{\textbf{Qwen2.5-7B}} 
     &\multicolumn{3}{c|}{\textbf{Qwen2-7B}} 
     &\multicolumn{3}{c}{\textbf{Qwen2.5-3B}} \\
    \cmidrule(r){2-4} \cmidrule(r){5-7} \cmidrule(r){8-10}
    &\textbf{MMLU}&\textbf{HumanEval}&\textbf{GSM8K} 
    &\textbf{MMLU}&\textbf{HumanEval}&\textbf{GSM8K} 
    &\textbf{MMLU}&\textbf{HumanEval}&\textbf{GSM8K} \\
    \midrule
    \textbf{API query}
    &147.80&62.15&57.92 &152.92&65.18&61.09 &92.46&38.89&44.91 \\
    \rowcolor{gray!15} \textbf{{\n}-Transfer}
    &165.60&70.10&69.80 &170.72&74.33&72.25 &111.57&52.35&50.34 \\
    \rowcolor{gray!15} \textbf{{\n}-SD}
    &269.35&103.84&130.89 &282.90&95.01&129.45 &173.85&90.30&85.34 \\
    \rowcolor{gray!15} \textbf{{\n}}
    &197.85&185.22&205.94 &215.68&199.62&218.53 &145.29&139.61&125.24 \\
    \bottomrule
  \end{tabular}}
\end{table*}

\begin{figure*}[]
    \centering  \includegraphics[width=6.8in]{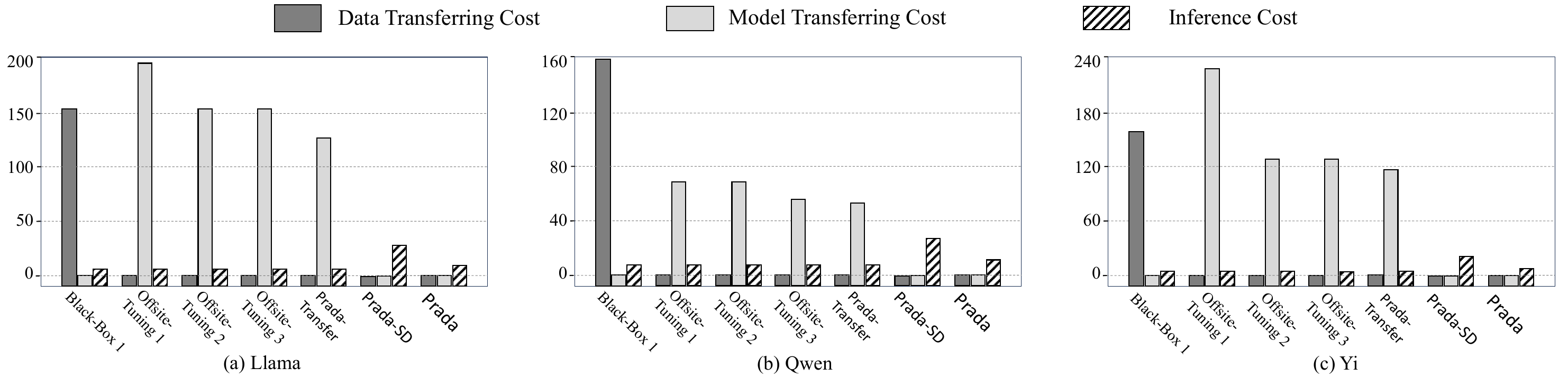}
    \vspace{-0.3cm}
    \caption{The communication costs of baselines and {\n}. The y-axis is the total communication cost (MB). The communication cost is calculated in three aspects: data transferring, model transferring, and inference cost. The data transferring is calculated by the Code-Instruction-120K training dataset. The inference stage cost is calculated by the communication of 1.5K queries.}
    \label{fig:commu}
\end{figure*}

\subsection{Fine-Tuning Cost Analysis}
A key advantage of {\n} is its use of lightweight proxy model instead of directly fine-tuning the large black-box LLMs, allowing device users to adjust local fine-tuning settings based on their resource constraints. Given that GPU memory is one of the most critical limitations on edge devices, it is essential to demonstrate that {\n} can significantly reduce peak memory usage during implementation. Here, we first evaluate the peak memory usage during the fine-tuning of {\n} and baselines.
Figure~\ref{fig:ftcost} presents comparisons of peak GPU memory usage (top row) and fine-tuning time (bottom row) for different methods across the Llama, Qwen, and Yi model families. The results clearly demonstrate that {\n} significantly reduces both memory consumption and fine-tuning time compared to baseline approaches.

\textbf{Peak GPU Memory Usage.} Standard SFT methods and Offsite-Tuning (solid gray bars) require substantially higher memory across all evaluated models. Specifically, direct SFT of large LLMs (solid gray bars) is practically infeasible on edge devices. For example, fine-tuning Llama-30b model exceeds 100GB of memory, far surpassing typical edge device capabilities. 
Offsite-Tuning (light gray bars) partially reduces memory demands by fine-tuning adapters rather than full models, but still requires substantial resources, such as 30GB of memory for Llama-13b. 
In contrast, {\n} (striped and dotted bars) drastically reduces peak memory usage. Proxy models used by {\n} consume less than 16GB in all scenarios. For example, using a small 1.5B proxy model (Qwen family), memory usage is further reduced to under 12GB, making {\n} feasible for most GPU-equipped edge devices. This substantial memory savings highlights {\n}’s practical advantage in resource-constrained environments.

\textbf{Fine-Tuning Time.} The total fine-tuning time follows a similar pattern.  Standard SFT methods require the longest fine-tuning time, particularly for larger models such as Yi-34B, where fine-tuning takes over 200 minutes. 
Offsite-Tuning reduces the training duration but still remains computationally demanding. 
In contrast, {\n} consistently achieves the shortest fine-tuning times across all model families. This improvement is especially notable in the Qwen and Yi families, where {\n} reduces training time by approximately 40–50\%. These significant reductions demonstrate that leveraging proxy models not only lowers GPU memory usage but also accelerates the fine-tuning process, making {\n} highly efficient in practice.

Overall, these results confirm that {\n} provides a computationally efficient and memory-friendly alternative to baselines, making it a practical solution for deploying privacy-preserving LLM adaptation on memory-limited edge devices.


\subsection{Communication Cost Analysis}
In addition to fine-tuning costs, practical deployment of {\n} on edge devices also depends heavily on communication efficiency. One major barrier to adapting black-box LLMs on edge devices is the substantial communication overhead associated with transmitting model updates. Standard SFT requires full weight updates, which involve transmitting large model checkpoints, often amounting to hundreds of megabytes or even gigabytes. Offsite-Tuning alleviates this burden by  transferring only adapter weights, but still necessitates frequent synchronization with the server, resulting in considerable communication overhead.

Figure~\ref{fig:commu} compares the communication costs for different adaptation methods across the Llama, Qwen, and Yi model families, breaking down the cost into \textit{data transfer}, \textit{model transfer}, and \textit{inference costs}. Fine-tuning the black-box LLM (Black-Box 1) incurs the highest data transfer cost due to the need for transmitting the entire training dataset. Offsite-Tuning avoids sending data but still requires transferring  adapter parameters, resulting in high model transfer costs, reaching nearly 200MB for Llama, 160MB for Qwen, and over 220MB for Yi. 
For {\n}, the communication cost is evaluated under three different strategies. {\n}-Transfer involves the client transmitting the fine-tuned LoRA modules to the server while performing inference entirely on the server. As shown in Figure~\ref{fig:commu}, {\n}-Transfer has a slightly lower model transfer cost than Offsite-Tuning due to the use of a smaller proxy model. In {\n}-SD, the proxy model remains on the client, and speculative decoding is employed to accelerate inference. In this case, there is no data or model transfer cost, though the inference communication cost increases slightly due to the need to exchange more tokens for speculative decoding. However, compared to the large model and dataset sizes, this additional inference cost is negligible and does not constitute a bottleneck for the system. {\n} without any optimization incurs the lowest overall communication cost. In conclusion, both {\n} and {\n}-SD significantly reduce communication costs by at least 80\% compared to baselines, demonstrating its feasibility for deployment in bandwidth-constrained environments such as edge devices. For applications where inference latency is critical, {\n}-Transfer offers a practical trade-off by moderately increasing communication to achieve lower latency.

\subsection{Token-wise Latency Analysis} 
Since {\n} adopts a client-server interaction-based generation method to preserve privacy and reduce communication overhead, it inevitably introduces additional communication costs during inference (as shown in Figure \ref{fig:commu}) and increases generation latency. However, in this section, we experimentally demonstrate that: (1) the additional latency introduced by {\n} is acceptable given its strong privacy guarantees, and (2) latency can be significantly mitigated using optimized strategies {\n}-Transfer and {\n}-SD.

Table~\ref{tab:latency} compares the token-wise generation latency of {\n} and baselines across different Qwen models and tasks. The latency is measured as the total runtime divided by the number of generated tokens. The \textbf{API query} method achieves the lowest latency across all scenarios, reaching as low as 38.89ms on HumanEval and 44.91ms on GSM8K for Qwen2.5-3B. The latency increases to 152.92ms on MMLU and 61.09ms on GSM8K for Qwen2.5-7B. 
Prada-Transfer introduces additional computation overhead but remains relatively efficient, increasing latency by only around 20-30ms/token compared to the API query.

In contrast, {\n}-SD and {\n} incur higher token-wise latency due to additional on-device processing overhead. {\n}-SD reaches 90.30ms on HumanEval and 85.34ms on GSM8K with Qwen2.5-3B, increasing further to 95.01ms and 129.45ms with Qwen2-7B. {\n} exhibits the highest latency, peaking at 139.61ms on HumanEval and 125.24ms on GSM8K for Qwen2.5-3B, and reaching 199.62ms and 218.53ms on Qwen2.5-7B. Notably, {\n}-SD demonstrates greater efficiency in long-form generation tasks such as HumanEval and GSM8K, reducing latency compared to {\n}. 

These results, together with the experimental findings of the communication cost, provide a comprehensive analysis of the trade-offs between communication cost and inference latency. 
For applications where \textbf{communication bandwidth} is the main limitation, {\n}-SD provides optimal efficiency, substantially reducing model transfer overhead while maintaining acceptable latency.
On the other hand, for latency-sensitive applications requiring \textbf{rapid responses}, {\n}-Transfer achieves near-API-level latency, though it involves moderate communication overhead.


\vspace{-0.2cm}

\subsection{Ablation Studies}
\begin{figure}[t]
    \centering  \includegraphics[width=3.4in]{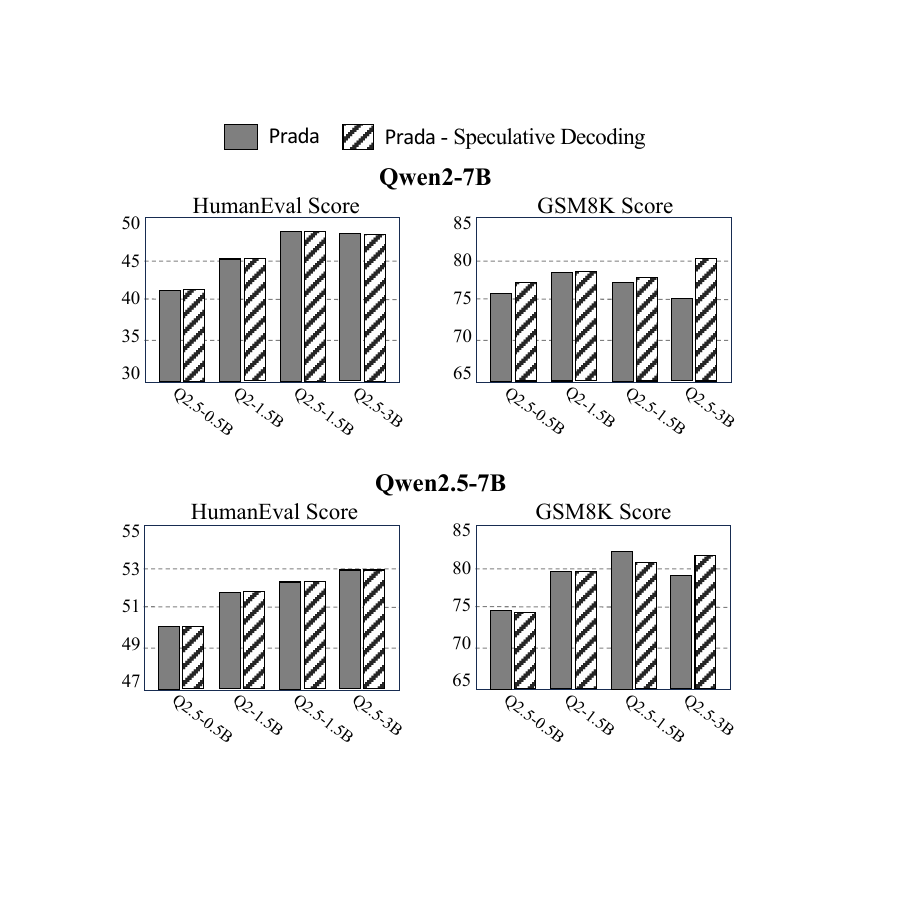}
    \vspace{-0.9cm}
    \caption{The impact of proxy model size and speculative decoding on HumanEval and GSM8K Scores of Qwen models.}
    \label{fig:ablation}
\end{figure}

In {\n}, the capacity of the proxy model affects the performance of {\n}. It is also crucial that most of the model sizes supported by edge devices work well for {\n}. Therefore, we compare the {\n} performance with different proxy model sizes in Figure \ref{fig:ablation}. In addition, theoretically, applying speculative decoding should not affect the system's performance. We still conducted experiments to verify this, and the results are presented as striped bars in Figure \ref{fig:ablation}.

\textbf{Impact of Proxy Model Size.} 
Figure \ref{fig:ablation} demonstrates that as the proxy model size increases, the performance of {\n} improves across both HumanEval and GSM8K benchmarks. Notably, transitioning from 0.5B to 1.5B yields significant gains, particularly on GSM8K, where performance stabilizes beyond this size. This suggests that while larger proxy models continue to enhance results, a 1.5B model is already sufficient to achieve strong adaptation, striking a balance between performance and computational efficiency.

\textbf{Impact of Speculative Decoding.} 
Figure \ref{fig:ablation} also indicates that speculative decoding has minimal impact on the overall performance of {\n}. Across all proxy model sizes, HumanEval pass@1 remains nearly unchanged, and GSM8K scores exhibit only marginal variations. This suggests that speculative decoding primarily serves as an acceleration technique without significantly altering model accuracy, making it an effective optimization strategy for reducing inference latency without compromising performance.


\section{Related Work}
\textbf{Privacy-Driven LLM Fine-Tuning.} 
Researchers have explored various strategies to facilitate fine-tuning while ensuring data and model privacy. FL~\cite{cheng2021fine,woisetschlager2024federated,ye2024openfedllm} enables collaborative model adaptation across multiple devices by allowing clients to perform local fine-tuning, with their model updates aggregated on a central server to obtain a globally optimized model. Recent advancements in FL for LLMs focus on improving LoRA aggregation~\cite{wang2024flora}, enhancing privacy preservation~\cite{sun2024improving}, and reducing computational and communication overhead~\cite{kuo2024federated,wang2025federated}.

Another key approach is Offsite-Tuning~\cite{xiao2023offsite}, which fine-tunes adapters of a compressed LLM while maintaining the privacy of the black-box model. Additional efforts in this direction leverage trusted execution environments~\cite{frikha2024obfuscatune} and dynamic layer replacement techniques~\cite{yao2024scaleot} to enhance privacy guarantees and efficiency. As concerns over model privacy continue to grow, black-box tuning methods have emerged that align with the objectives of {\n}. For instance, \cite{sun2022black} proposed a black-box tuning framework that optimizes model performance using only API access to the LLM, though it primarily focuses on prompt tuning rather than full model adaptation. Similarly, \cite{li2024federated} integrates prompt tuning with FL, further strengthening data privacy protection.

\textbf{Parameter-Efficient LLM Fine-Tuning.}  
A wide range of PEFT techniques have been developed to reduce memory overhead in LLM adaptation. Adapter-based methods~\cite{houlsby2019parameter,he2021towards,zhu2021counter} introduce trainable modules within the model, while subsequent work~\cite{pfeiffer2020adapterfusion,wang2022adamix} improves upon this by exploring adapter fusion and mixing strategies. Another major category is low-rank adaptation (LoRA)~\cite{hu2021lora}, which models weight updates using low-rank matrices, substantially lowering memory consumption. Several works~\cite{liu2024dora,kopiczko2023vera} further refine LoRA through optimized low-rank decomposition, improving both efficiency and performance. 
Dynamic rank-based PEFT methods~\cite{zhang2023adalora,zhang2024autolora} adjust rank allocations during fine-tuning to achieve better adaptation efficiency. In contrast, sparse parameter tuning techniques, such as BitFit~\cite{zaken2021bitfit,lawton2023neural} and Diff Pruning~\cite{guo2020parameter}, selectively fine-tune only a subset of model parameters, further reducing memory and computation requirements. Collectively, these methods offer effective strategies for fine-tuning large-scale models with constrained resources.

\vspace{-0.1cm}

\section{Discussion}
This paper conducts a detailed discussion and optimization of user-server interactions and resource utilization in offset adaptation. Additionally, we investigate the impact of model size and the efficient use of system resources through experiments. Nonetheless, significant opportunities remain for future exploration. In particular, algorithmic improvements could focus on adaptive optimization of the offset and the feasibility of using distilled models as proxy models to further enhance system performance.

Another promising research direction building on {\n} involves enhancing prompt privacy. Although our system effectively protects the privacy of training datasets and proprietary model parameters, user prompts must still be transmitted to the server during inference. These prompts could contain sensitive or private information, raising important privacy concerns. Thus, a key challenge is how to safeguard prompt privacy while maintaining effective inference. Potential solutions include leveraging trusted third-party computing platforms~\cite{luo2023bc4llm}, which could further enhance data privacy protection and provide a more comprehensive system.

For scenarios in which token-wise latency represents the main performance bottleneck, we introduced the {\n}-Transfer strategy, wherein the adapted proxy model is transmitted to the server. While Offsite-Tuning follows a similar model-transfer approach, it introduces additional privacy vulnerabilities, as servers could potentially reconstruct sensitive data from model updates using model inversion techniques~\cite{fredrikson2015model,zhang2024extracting}. To address these risks, future work could incorporate defenses against model inversion attacks, such as model obfuscation or privacy-preserving transformations~\cite{gong2023netguard}, or again leverage secure third-party platforms for inference.

\section{Conclusion}
In this work, we introduced {\n}, a privacy-preserving system designed for black-box LLM adaptation on resource-constrained edge devices. By utilizing a lightweight proxy model fine-tuned locally with LoRA, and employing adaptation offsets to enhance inference from a remote black-box LLM, {\n} effectively addresses critical challenges related to data and model privacy. Through comprehensive experiments, we demonstrated that {\n} achieves substantial efficiency improvements, significantly reduces communication overhead, and maintains competitive performance across multiple evaluation benchmarks.
We also performed an in-depth analysis of system resource utilization and explored how model size impacts adaptation effectiveness. By optimizing client-server interactions and computational efficiency, {\n} presents a practical solution for real-world deployments where both privacy and performance are essential. Furthermore, we evaluated enhancements such as speculative decoding, showcasing their beneficial effects on inference latency and overall system usability.

Despite these contributions, several open challenges remain. A key direction for future work is improving the adaptability of offset-based optimization techniques across different LLM architectures. Additionally, exploring the use of distilled models as proxy models may further enhance system efficiency. Another critical issue is ensuring prompt privacy, as prompts may still contain sensitive information that must be transmitted to the server. Techniques such as secure computation and trusted execution environments could further strengthen privacy guarantees and provide a more comprehensive framework for private LLM adaptation.

Overall, {\n} lays the foundation for privacy-aware and efficient black-box LLM adaptation, opening a promising direction for future research in privacy-preserving model adaptation and accelerated inference on edge devices.

\bibliographystyle{ACM-Reference-Format}
\bibliography{referen}

\appendix

\end{document}